\documentclass[letterpaper,11pt]{article}

\usepackage{epsfig}
\usepackage{amsfonts}
\usepackage{graphics}
\usepackage{amssymb}
\usepackage{natbib}
\bibpunct{(}{)}{;}{a}{,}{,}

\newcommand{\be}{\begin{equation}}
\newcommand{\ee}{\end{equation}}
\newcommand{\bea}{\begin{eqnarray}}
\newcommand{\eea}{\end{eqnarray}}

\def \Re{{\rm I\kern -1.6pt{\rm R}}}
\def \Expect{{\rm I\kern -1.6pt{\rm E}}}

\begin{document}
\bibliographystyle{plainnat}


\twocolumn[%
\begin{@twocolumnfalse}
\begin{center}{\Large \bf Identifying statistically significant patterns in
  gene expression data} \\ \vspace{5mm}
 {\Large Patrick E. McSharry$^{1}$ and Edmund J. Crampin$^{2,*}$}
\end{center}

\begin{center}
$^1$ Oxford Centre for Industrial and Applied Mathematics,
  Mathematical Institute, Oxford, OX1 3LB, UK and Department of Engineering
  Science, University of Oxford, Oxford, OX1 3PJ, UK, and  $^2$ Centre for
  Mathematical Biology, Mathematical Institute, Oxford, OX1 3LB, UK,
  University Laboratory of Physiology, University of Oxford, Oxford, OX1 3PT.
\end{center}

\noindent
$*$ To whom correspondence should be addressed. \\
Both authors contributed equally to this work. \\

\noindent
Manuscript Date: 9th August 2002
\vspace{5mm}

\noindent
{\em Running head:} Statistical significance in gene expression data
\vspace{5mm}

\end{@twocolumnfalse}
]

\begin{abstract}
\subsection{Motivation}
Clustering techniques are routinely applied to identify patterns of
co-expression in gene expression data. Co-regulation, and involvement 
of genes in similar cellular function, is subsequently inferred from the
clusters which are obtained. 
Increasingly sophisticated algorithms have been applied to microarray data,
however, less attention has been given to the statistical significance of the
results of clustering studies. 
We present a technique for the analysis of commonly used hierarchical
linkage-based clustering called Significance Analysis of Linkage Trees (SALT).
\subsection{Results}
The statistical significance of pairwise similarity levels between gene
expression profiles, a measure of co-expression, is established using a
surrogate data analysis method. We find that a modified version of the
standard linkage technique, {\em complete-linkage}, must be used to generate
hierarchical linkage trees with the appropriate properties. 
The approach is illustrated using synthetic data generated
from a novel model of gene expression profiles and is then applied to
previously analysed microarray data on the transcriptional response of human
fibroblasts to serum stimulation. 
\subsection{Availability}
A set of MATLAB functions are available on request. 
\subsection{Contact}
edmund.crampin@unimelb.edu.au
\end{abstract}

\section{Introduction}
The ability to measure expression levels of multiple genes simultaneously
promises insights into the regulation of gene expression under
both normal and pathological conditions. Gene expression data is now
routinely collected using oligonucleotide and cDNA microarray
technologies \citep{fodor:1993a,pease:1994a,schena:1996a}.
Microarray experiments
involve many separate steps in the preparation of samples, of the
arrays themselves and in the subsequent image acquisition and analysis
\citep{hauser:1998a}.
However, as the reliability of the data
improves, there is an increasing need for tools for data analysis and
interpretation. While increasingly sophisticated clustering techniques
are being applied to microarray data, the statistical significance of
clustering results has yet to be fully explored.

A wide variety of statistical techniques
have been used to investigate these gene expression profiles,
including principal component analysis
\citep{holter:2000a,alter:2000a,holter:2001a}, correspondence analysis
\citep{fellenberg:2001a}, neural networks~\citep{herrero:2001a} and the
construction of statistical models \citep{zhao:2001a,ramoni:2002a}. 
A common starting point for the analysis of microarray
data is to use a clustering technique to
group together genes with similar expression profiles
\citep{wen:1998a,eisen:1998a,yeung:2001a}.  
Genes exhibiting similar patterns suggest co-regulation of gene
expression, and co-expressed genes may be involved in similar functions
within the cell. 
The ultimate goal of such studies is to be able to predict the underlying
gene networks giving rise to the gene expression data \citep{yeung:2002a}.


In this paper we aim to determine the significance of the similarity between
gene expression profiles. We deal with data collected over several time
intervals (time series data), although the technique applies equally for gene
expression data recorded over multiple separate experiments. In particular, we
show how the significance of the number of clusters emerging from
linkage analysis techniques can be assessed. Linkage analysis, a simple and
widely used clustering technique, performs 
clustering by sorting the gene expression profiles according to
pairwise similarity. A distance metric is used to quantify the similarity
between genes, where the closer two genes are in distance the more similar
are their expression patterns, i.e. the more likely they are to be
co-expressed. 
The ordered gene expression profiles can be represented graphically using a
tree (called a dendrogram) where the position of the branch connecting two
genes reflects the similarity of their expression profiles. Sorting the
genes in this way results in similar expression profiles being grouped
together. 
Currently many investigators identify clusters of interest from the tree by
both visual inspection and {\it a priori} knowledge. 
The tree identifies many different possible numbers
of clusters, ranging from one extreme where there are as many clusters
as genes (one per cluster) to the other extreme, where one cluster
contains all genes in the data set. Cutting the tree at a particular
distance value establishes a number of clusters into which the
expression profiles are grouped. 

The major difficulties in the analysis of microarray data, as for most data
sets, arise in discriminating between signal and noise. Given limited data,
there will always be a nonzero probability of incorrectly identifying noise
as genuine co-expression. For microarray data the situation is exacerbated by
the generally low number of observations made on large numbers of variables
(genes). Any statistical test applied to the data must indicate an acceptable
level of such `false positives' arising. Statistical tests rely on
assumptions made about what constitutes noise in the data set, against which
a null hypothesis is tested. The nominal level of the statistical test will
only be meaningful if the assumptions about the noise are correct, and the
null hypothesis relevant to the data set. We use the
method of surrogate data \citep{theiler:1992a,smith:1992a} to determine a
threshold on the tree at which we can reject, with a
given confidence level, the null hypothesis that observed similarity values 
could have occurred by chance. This threshold can be used to determine which
clusters are significant. 

We demonstrate the approach (Significance Analysis of Linkage Trees, SALT)
using synthetic gene expression data
generated using a simple model, and subsequently apply the surrogate
analysis technique to a publicly available gene expression data set
which analyses the response of human fibroblasts to serum
\citep{iyer:1999a}.

\section{Methods}
Gene expression data is commonly expressed as the logarithm of the
ratio of an observed signal to the initial or other reference
expression level. Hence values are initially zero, and a positive
value indicates up-regulation, whereas negative values represent
down-regulation of gene expression, and up- and down-regulation are
given equal numerical importance. 

\subsection{Synthetic Gene Expression Data}
\begin{figure}[h]
\centerline{\psfig{file=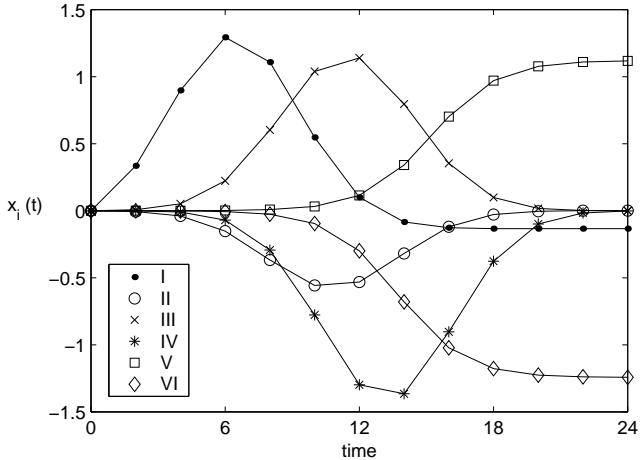,width=86mm}}
\caption{Examples of the six response functions used for generating synthetic
  expression data (with no observational noise added).}  
\label{f:genesynthetic}
\end{figure}

To illustrate how linkage clustering performs on a data set
containing predetermined patterns of co-expression of genes, where gene
clusters are known, we generated synthetic gene
expression data. Six time-dependent response functions (labelled I to VI)
were used to simulate early and late response, and up- and down-regulation
expression patterns (Fig. \ref{f:genesynthetic}). 
We used a Gaussian function 
$ f_g(t,\tau,c) = \\ \exp [ - \frac{1}{2} (\frac{t - \tau_r - \tau}{c})^2]$
to simulate genes switching on and off transiently during the experiment 
(response functions I through IV in Fig. \ref{f:genesynthetic}) and a
sigmoid function 
$f_s(t) = \frac{1}{2}[ 1 + {\rm tanh} (\frac{t-\tau_r-\tau}{c})] $
to replicate genes responding slowly during the experiment, to reach a
new threshold expression level (response functions V and VI in
Fig. \ref{f:genesynthetic}). Here $\tau_r$ is a time delay for each of the six
responses, and $\tau$ and $c$ are gene-specific delays and timescales. 

A time series of $T=13$ points was used to mimic the change in gene
expression data over a 24hr period. Our synthetic data set contained $N =
120$ genes: $20$ genes for each of the six responses.
Co-expressed genes may respond after different delays and over
different timescales, and will show different amplitudes of expression. 
We incorporated such variation into the amplitudes of the gene expression
profiles and the response times by representing the expression level for
gene $i$ through 
$x_i(t) = A_i f(t,\tau_i,c_i) + \eta_i(t)$,
where $f$ stands for either up- or down-regulation with response function
$f_g$ or $f_s$, as described in Table \ref{t:param}. 
For each gene $i$, the variation in the amplitude $A_i$, time delay
$\tau_i$ and timescale $c_i$ are sampled from uniform distributions
with $A_i \in [0.5,1.5]$, $\tau_i \in [-1,1]$ and $c_i \in [2.9,3.1]$.
In addition, we simulate observational uncertainty due to measurement errors
by adding a normally distributed noise term $\eta_i(t)$ with zero mean and
standard deviation 0.05.
\begin{table}[ht]
\caption{Parameters for generating the synthetic data set.} 
\begin{center}
\begin{tabular}{ccccccc}
\hline
Response $r$  &I    &II    &III    &IV   &V  &VI \\
\hline
$f$   &$+f_g$  &$-f_g$  &$+f_g$  &$-f_g$  &$+f_s$  &$-f_s$ \\
$\tau_r$  &7  &10  &12  &14  &15  &14\\
\hline
\end{tabular}
\label{t:param}
\end{center}
\end{table}

\subsection{Fibroblast Data Set}
We used the published data set of \citet{iyer:1999a}
for the response of human fibroblasts to serum following serum
starvation. Data was collected at 12 times over a 24hr period for
around 8,600 distinct human transcripts. A further data point is
included for exponentially growing cells (``unsynch'') to give a
series of $T=13$ observations. Of these genes 517 were found to change
expression levels in response to serum stimulation (for more
information see \citet{iyer:1999a} and the
accompanying website http://genome-www.stanford.edu/serum). 
We restricted our analysis to these 517 genes.

\subsection{Similarity Measure}
In order to cluster a data set containing gene expression profiles
using a linkage algorithm, a mathematical
definition of the similarity between expressions is required.
Two genes which are
co-expressed are likely to be similar in shape, but not necessarily in
magnitude and for this reason the correlation coefficient is a
suitable similarity metric.  Following \citet{eisen:1998a}, denoting the 
logarithm of the expression ratio for gene $i$ at time $k$ by $x_{ik}$, 
the similarity between genes $i$ and $j$ is quantified by
$\rho_{ij} = \frac{1}{T} \sum_{k=1}^T ( \frac{x_{ik}}{\sigma_i} )
( \frac{x_{jk}}{\sigma_j} )$,
where
$\sigma_i = \sqrt{ \frac{1}{T} \sum_{k=1}^T x_{ik}^2 }$
and $T$ is the number of observations.
Note that if the means were subtracted from each of the expression profiles 
then $\sigma_i$ would be the standard deviation and $\rho_{ij}$ the
correlation coefficient \citep{chatfield:1989a}.
By explicitly setting the mean to zero \citep{eisen:1998a}, 
corresponding to an expression ratio of one, we are selecting a 
reference state against which subsequent changes are contrasted.
Values of $\rho_{ij}$ can vary from $1$ (completely correlated) to $-1$
(completely anti-correlated), whereas $\rho_{ij} = 0$ implies that the
two genes are uncorrelated.

The similarity measure $\rho_{ij}$ takes negative values for
anti-correlated
data and so cannot be used as a distance between two observations. A
distance measure $d_{ij}$ can be calculated from $\rho_{ij}$ using
$d_{ij} = \sqrt{2(1-\rho_{ij})}$
which fulfils the conditions required for a distance
metric: namely (i) $d_{ij} = 0$ if and only if $i=j$, i.e. the genes have
the
same expression profiles, (ii) $d_{ij} = d_{ji}$ and (iii) $d_{ij}
\leq d_{ik} + d_{kj}$ \citep{mantegna:1999a}. 
The $N(N-1)/2$ unique distances $d_{ij}$ ($i=1,\ldots,N$,
$j=i+1,\ldots,N$)
can be used to determine a tree connecting $N$ genes using a graph
consisting of $N-1$ linkage distances. Clearly, this reduction in the 
number of distances used to generate the tree implies a loss of information. 
The aggregation process for constructing the tree must be carefully chosen 
so that important information is not lost, and can be achieved by using a 
suitable linkage algorithm.

\subsection{Clustering by Complete-Linkage}
Linkage algorithms iteratively combine the $N$ genes into clusters.
A measure of affinity between clusters is used to decide the order in
which clusters are combined at any given step.  Starting from $N$
clusters, each consisting of a single gene, the two clusters with the
highest affinity are combined into a new cluster. This linkage is
marked on the tree by a connection between the clusters
at a linkage distance equal to the affinity value.  This process is
repeated until there is only one large cluster containing all $N$
genes.  If the affinity between clusters is chosen to be the distance
between the closest pair of genes then the method is known as
single-linkage. The tree constructed with single-linkage
is called a minimum spanning tree. This has the disadvantage that 
the linkage distance does not place a bound on how dissimilar genes within
the same cluster may be. This is because the distances $d_{ij}$ for each of
the other pairings of genes between the clusters will be
larger than the linkage distance, $\delta_{ij}$, by construction. 
Choosing the distance between the averages of the
clusters as the measure of affinity is known as average-linkage, and
is a choice commonly used in microarray data analysis 
\citep{eisen:1998a}. Similarly, for average-linkage $\delta_{ij}$ does not
place an upper bound on the pairwise distances within clusters.

We suggest that the appropriate linkage technique to choose is
complete-linkage, where the maximal pairwise distance between
clusters $\delta_{ij}$ is used to determine the tree. At each step
the two clusters with the smallest maximal pairwise distance are
combined. Thus the
tree contains maximal distances between clusters, and hence is
appropriate for an analysis of statistical significance.
In particular, for the complete-linkage algorithm all genes clustered below a
threshold distance $d_{ij}=d_{thrs}$ must be separated by distances 
satisfying $d_{ij} \le d_{thrs}$. If we cut the tree at $d_{thrs}$ and ignore
all linkages above $d_{thrs}$ 
then we are certain not to neglect any gene pairings which are separated by a
distance $d_{ij} \le d_{thrs}$. This property of the complete-linkage
algorithm will be employed to obtain significant clusters. 



It still remains to be decided at which distance to cut the tree, to
determine the number of clusters which with high probability represent
co-expression of their constituent genes.
To obtain the distribution of distances that one would expect to find for
genes which are not being co-expressed, we have generated surrogate data
\citep{theiler:1992a,smith:1992a}.

\subsection{Surrogate Gene Profiles}
The SALT technique determines how small the distance between two genes must
be before they are inferred to be co-expressed. 
This is achieved by testing against a particular null hypothesis: 
that a particular value of the distance arises by chance from two genes 
which are not co-expressed.  
We test this null hypothesis using the distribution of distances which 
results from genes that are not co-expressed. 
Since we do not have an analytical description for this distribution, 
we estimate it using surrogates gene expression profiles, generated by 
sampling from the original data set.
A significance level must be specified at which the null hypothesis is
tested.  For example, if we allow a 5\% chance that
the null hypothesis is falsely rejected, then the test is valid at the 95\%
level. 
Testing at the 95\% significance level means that we reject the null
hypothesis for any distance in the tree less than a threshold distance 
$d_{thrs}$ corresponding to the 5th percentile of the distribution 
obtained from the surrogates. 
Each pair of surrogate gene expression profiles provides one sample 
of the distribution.  A large number of surrogates (10000) was used to 
resolve the tail of the distribution, giving a robust estimate of 
the 5th percentile. 

\begin{figure}[t]
\centerline{\psfig{file=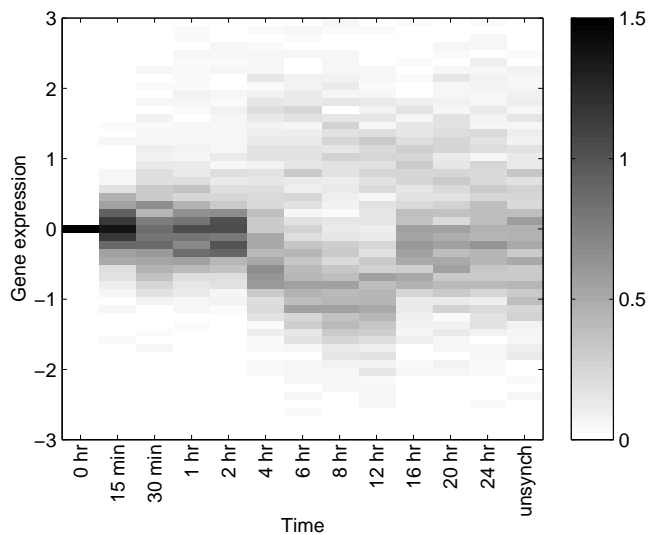,width=86mm}}
\caption{Probability density function (PDF) of the fibroblast gene
expression data set.} 
\label{f:serumpdf}
\end{figure}

We generate surrogate expression profiles which preserve some aspects of 
the original data but which are also consistent with the null hypothesis. 
Appropriate surrogates should reflect obvious properties of the data, in
particular that the gene expression profiles vary smoothly with time.
An analysis of the probability density function (PDF) of the gene
expression profiles for the fibroblast data set (the probability that, at
a given time point, a gene selected at random from the data set will
have a given expression level) shows that there is a time-dependent
trend running through the data (Fig. \ref{f:serumpdf}).  This suggests
that surrogate data sets based on permutations which shuffle the
temporal information are likely to destroy important correlations
which are due to the time series nature of the experiment.
To preserve the temporal continuity we have constructed surrogate gene
expression profiles by sampling without replacement from the original 
data at each time point independently. 
In this way the PDF of the gene expression at different
time points (Fig. \ref{f:serumpdf}) is preserved in the surrogates. 

We can then reject the null hypothesis for distances calculated from the
original data for which $d_{ij}<d_{thrs}$ with confidence 95\%. The
construction of the tree using complete-linkage algorithm ensures that 
for clusters below $d_{thrs}$, the distance between every gene pair 
within a cluster is statistically significant. Note that this is not the case
for the other linkage algorithms described above.

\section{Implementation}
We applied complete-linkage clustering to both the synthetic and
fibroblast data sets.
Surrogates were generated to find $d_{thrs}$ and this
threshold was used to determine statistically significant clusters. We
found that 10000 surrogates was sufficient to produce robust
results when testing at the 95\% significance level.

\subsection{Synthetic data}

\begin{figure*}[t]
\centerline{\psfig{file=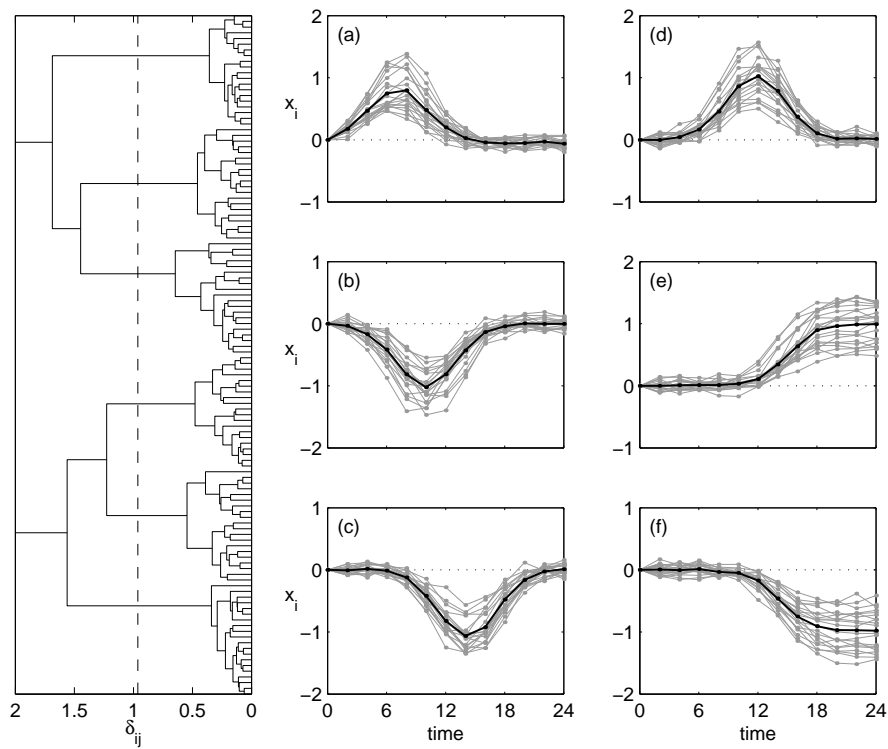,height=10cm}}
\caption{Analysis of the synthetic data set showing 
the hierarchical tree (left) constructed using complete-linkage, with distances
$\delta_{ij}$ from 2 (no co-expression) to 0 (co-expression). The clusters
recovered by cutting the tree at significance threshold $d_{thrs}=0.963$
(indicated on the tree by the dashed line) are shown on the right. The darker
trace indicates the cluster average expression profile.} 
\label{f:synthresults}
\end{figure*}

The distances corresponding to the synthetic data set
(Fig. \ref{f:synthresults}) fall into distinct groups because of the
clearly defined response profiles underlying the data
(Fig. \ref{f:genesynthetic}). 
The tree (left) shows the hierarchical
organisation of the linkage distances, indicating that the correct number of
clusters can be identified by cutting the tree at a threshold distance in the
range $0.646 < d_{ij} < 1.227 $.  
Surrogate analysis provided a threshold distance of $d_{thrs}=0.963$
corresponding to the 95\% significance level. This threshold correctly
identifies the clusters, assigning the 20 genes for each response
function to the corresponding cluster. The correspondence between the original
responses 
(Fig. \ref{f:genesynthetic}) and the clusters (Fig. \ref{f:synthresults}) is
given by: I $\mapsto$ (a), II $\mapsto$ (b), III $\mapsto$ (d), IV $\mapsto$
(c), V $\mapsto$ (e) and VI  $\mapsto$ (f). 
Note that the expression profiles are clustered together, even though their
magnitudes are different, because they have the same shape, as quantified by
the choice of distance metric based on the correlation coefficient. 

We expect that significantly higher noise levels will obscure the
original pattern of gene expression. Increasing the observational noise in
the synthetic data set was found to increase the linkage distances used to
construct the tree, but not to have a strong influence on the threshold
distance (data not shown). 
Many more of the distances in the data set are found to be consistent with
the null hypothesis, and therefore not considered significant for 
clustering. In this case, a larger number of clusters than the six original
responses is obtained by cutting the tree at the threshold distance, however,
the members of each cluster were still found to correspond to only one of the
six response functions. 

\subsection{Fibroblast Data Results}
We applied the same surrogate data analysis technique to the 517 gene
expression profiles in the published data set corresponding to
fibroblast transcriptional response to serum stimulation. The published
clustering for this data set found clusters
corresponding to different aspects of the physiological response of
fibroblasts to wound healing (serum stimulation).
Iyer et al. used hierarchical clustering 
\citep[using average-linkage,][]{eisen:1998a} from which they obtained
clusters by visual 
inspection of the ordered tree, reporting cluster sizes of 142,
100, 60, 40, 32 and 31 genes for the six largest clusters. 

\begin{figure*}[htp]
\vspace{-1cm}
\centerline{\psfig{file=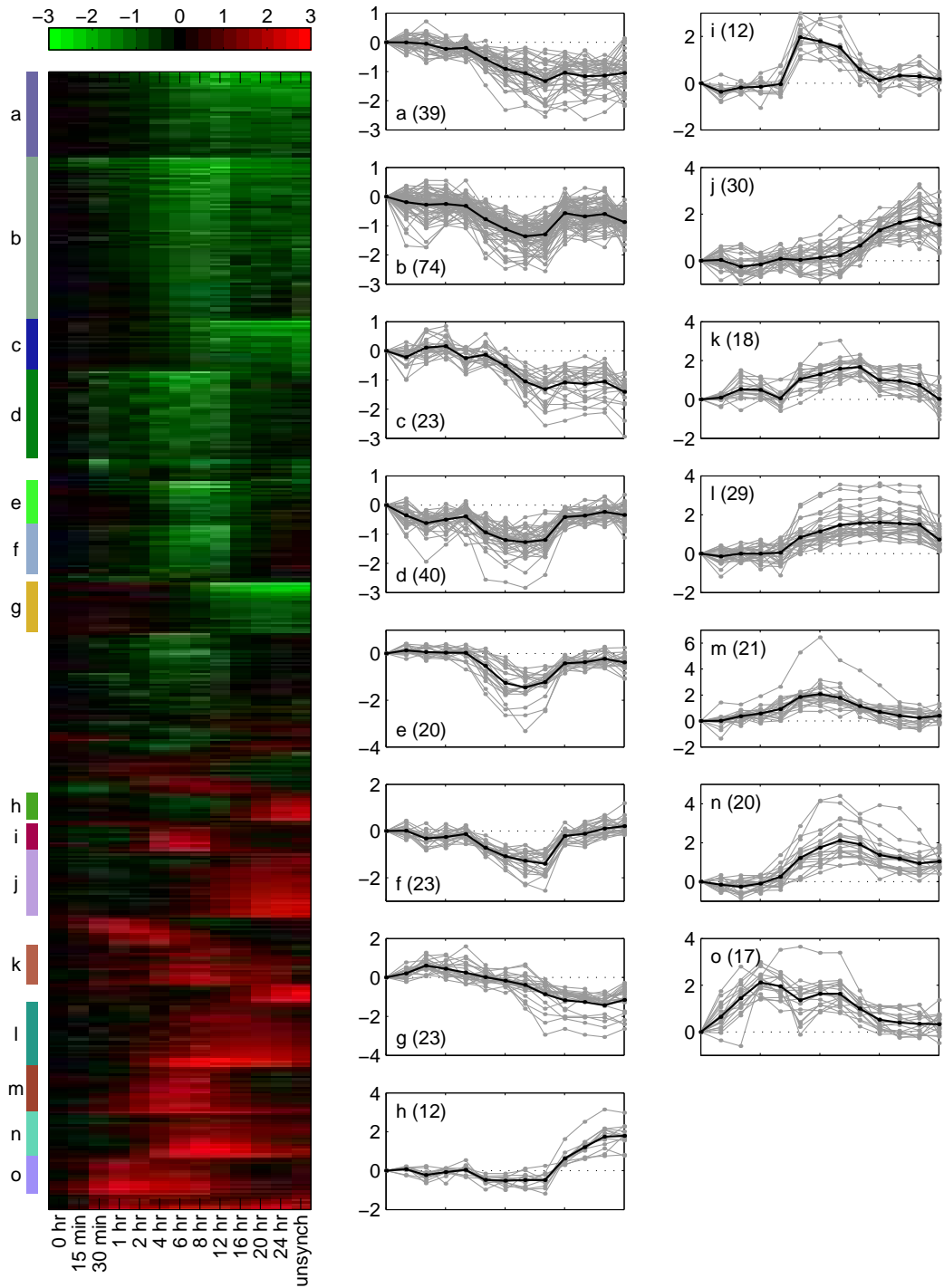,height=19cm}}
\caption{Analysis of the fibroblast gene expression data. The
  red-green display shows the clustered data set sorted by
  increasing mean expression level. The scale is log-expression ratio $-3$ 
  (down-regulation by a factor of 8) to $+ 3$ (up-regulation by a factor
  of 8). The 15 largest
  clusters obtained for significance threshold $d_{thrs}=0.95$ correspond
  to the locations indicated by coloured bars left of the red-green display 
 (cluster sizes are shown in brackets).}
\label{f:serumresults}
\end{figure*}

We used surrogates to obtain a threshold
distance of $d_{thrs} = 0.95$ corresponding to the 95\%
significance level, which was used to cut the tree.
The fifteen largest clusters obtained, those
containing 10 or more genes, are shown in Fig. \ref{f:serumresults}, along
with the red-green display \citep{eisen:1998a} of the entire
clustered data set, ordered by increasing mean expression value. 

In total 47 statistically significant clusters were found, including
30 containing more than two genes. 
The maximum cluster size found to be of statistical significance by
this technique is 74, and the six largest clusters are of size 74 (b), 40
(d), 39 (a), 30 (j), 29 (l) and 23 (c,f,g), suggesting that some of the
correlations  within the larger clusters reported previously may not be
statistically significant.

\section{Discussion}
The measurement of gene expression data using DNA microarray techniques
may provide a deeper understanding of many of the complicated
processes underlying biological systems. To investigate the large data sets
resulting from microarray experiments we have presented a new technique
(SALT).
This technique is straightforward and can be easily implemented in
existing computer analysis packages.

SALT uses surrogate data analysis to identify clusters on the hierarchical
tree generated using the complete-linkage algorithm, at a prescribed
significance level. 
We have used surrogate data which preserves the temporal
variation of the gene expression profiles. This is also the appropriate
method for data collected for different experimental conditions or from
different tissue preparations, in order to preserve characteristics of the
data within the different experiments. A better understanding of the
dynamical processes underlying the gene expression data and, in particular,
the sources of measurement error, would allow surrogates to be constructed
using better models of the properties of independent gene expression data.

The application of the technique to the published fibroblast data set found
significant clusters containing fewer genes than previous analyses
\citep{ramoni:2002a,iyer:1999a}. One explanation is that previous 
analyses employ similarity values between data points which are not
statistically significant to obtain their
clusters. Studies using alternative methods of clustering
in which the data is divided into a predetermined number of
clusters (K-means clustering, for example) also risk grouping together
genes for which the expression profiles have distances above the significance
threshold. 

Statistical significance testing using SALT could also be used for
clustering other examples of large data sets where one wishes to
identify which elements of a given set of profiles are (i) interacting
with each other or (ii) interacting in a similar way in response to
some external perturbation. 

\section*{Acknowledgements}
This research was supported by an Engineering and Physical Sciences 
Research Council Grant GR/N02641 to PEM and The Wellcome Trust (EJC).

{\small
\bibliography{dna}}
\end{document}